\documentclass[twocolumn,showpacs,preprintnumbers,amsmath,amssymb]{revtex4}
\usepackage{graphicx}
\preprint{LSGMO/2007}
\begin{document}
\DeclareGraphicsExtensions{.eps, .jpg}
\bibliographystyle{prsty}
\input epsf

\title{Effect of Mn substitution by Ga on the optical properties of a metallic manganite}
\author{A. Nucara$^{1}$, P. Maselli$^{1}$, M. Del Bufalo$^{1}$, M. Cestelli Guidi$^{2}$, J. Garcia$^{3}$, P. Orgiani$^{4}$, L. Maritato$^{4}$, and P. Calvani$^{1}$} 
\affiliation {$^1$"Coherentia" CNR-INFM and Dipartimento di Fisica,
Universita' di Roma ``La Sapienza'', Piazzale A. Moro 2, I-00185 Roma, Italy}
\affiliation {$^{2}$Laboratori Nazionali INFN di Frascati, Via E. Fermi, Frascati} 
\affiliation {$^{3}$Instituto de Ciencia de Materiales de Aragon and Departemento de Fisica de la Materia Condensada, Consejo Superior de Investigationes Cientificas y Universidad de Zaragoza, 50009 Zaragoza, Spain}
\affiliation {$^{4}$"Coherentia" CNR-INFM and Dipartimento di Matematica ed Informatica, Universit\'a di Salerno, Salerno, Italy}
\date{\today}

\begin{abstract} 
In a metallic manganite like La$_{2/3}$Sr$_{1/3}$MnO$_3$, the substitution of Mn$^{+3}$ by Ga$^{+3}$ dilutes the ferromagnetic order and locally cancels the Jahn-Teller distortion, without heavily affecting the crystal structure. One can thus follow the changes in the charge dynamics induced by Ga, until the  ferro-metallic manganite is turned into an insulator. Here this phenomenon is studied in detail through the infrared reflectivity of five samples of La$_{2/3}$Sr$_{1/3}$Mn$_{1-x}$Ga$_x$O$_3$,  with $x$ increasing from 0 to 0.30 and for  $50 \leq T \leq 320$ K. A simple model which links the measured optical parameters to the magnetization $M(x, T)$ well describes the behavior of the plasma frequency, the scattering rate, and the mid-infrared absorption along the metal-to-insulator transition.
\end{abstract}


\maketitle

\section{Introduction}

The close interplay between charge dynamics and magnetism is a basic feature of the manganites A$_{1-x}$B$_{x}$MnO$_{3}$ (A = La,Nd,Bi,Ce; B = Sr,Ca), as well as the starting point for any description of their rich phase diagram. It originates from the mixed valence of Mn ions, which can transfer both charge and spin between their states Mn$^{+3}$  and Mn$^{+4}$.
As a result, in many manganites the ferromagnetic (FM) phase coincides with a metallic state and is characterized by a negative magnetoresistance. Long time ago, this phenomenon was successfully explained in terms of the Mn$^{+3}$ - O$^{-2}$ - Mn$^{+4}$ charge-transfer mechanism, called double-exchange \cite{Zener,Anderson}. Due to the Hund's rule, this transfer occurs only if the core spin of Mn$^{+3}$ is aligned with that of Mn$^{+4}$, namely if the electron travels in a ferromagnetic environment. Otherwise, a Hund's energy of about 2 eV must be paid, as for example one observes in a manganite which switches at the N\'eel temperature from an FM to an antiferromagnetic (AF) phase \cite{Nucara06}. 

However, the double-exchange mechanism is found to dominate only in a narrow range of hole doping $y$. In fact, the $y,T$ phase diagram of these compounds also includes paramagnetic, FM insulating, AF, as well as charge- and orbital-ordered phases\cite{Cheong}. Then, the original double-exchange model has been improved in the years by introducing superexchange, correlation, and polaronic\cite{Millis} effects. The formation of charge polarons in manganites is due to the Jahn-Teller distortion that the oxygen octahedra around the Mn$^{4+}$ ions experience as they become Mn$^{+3}$ by receiving an itinerant electron. 
Even if this purely ionic picture is probably oversimplified, and both the electronic states and the Jahn-Teller distortions in mixed valence manganites are shared by different Mn ions\cite{Subias}, this approach successfully explains their optical spectra\cite{Noh,Ahn}. Therefore, it will be substantially adopted also in the present work. 

In the above framework, the controlled substitution of the mixed-valence Mn$^{+3}$ - Mn$^{+4}$ ions by other elements has been considered for many decades as a powerful tool to understand in further detail the role of magnetic and translational order in the electrodynamics of manganites. Since the pioneering work of Goodenough and coworkers\cite{Good1}, oxides where Mn was replaced by Al, Sc, Ga, Co, Ni, etc., have been obtained and studied by a variety of techniques\cite{Good2,Coldea,Blasco02,Farrell}. Ga$^{+3}$ is one of the most suitable ions to that purpose. Indeed, being its ionic radius (0.062 nm) quite close to that of Mn$^{+3}$ (0.0645 nm), Ga$^{+3}$ replaces Mn$^{+3}$ by leaving the manganite lattice only slightly perturbed\cite{Shannon} and the Mn$^{+4}$ concentration unaltered. However, the effect of Ga on the properties of an FM manganite is dramatic, as it has neither a magnetic moment, nor $e_g$ electrons in the 3d orbital which may produce Jahn-Teller distortions. Both in\cite{Yusuf,Sun} La$_{2/3}$Ca$_{1/3}$Mn$_{1-x}$Ga$_x$O$_3$ and in\cite{Blasco03} La$_{2/3}$Sr$_{1/3}$Mn$_{1-x}$Ga$_x$O$_3$, for increasing $x$, the Curie temperature $T_c$ decreases rapidly. The insulator-to-metal transition temperature $T_{IM}$ drops even more sharply. Both these effects have been attributed to the magnetic and topological disorder introduced by Ga, as well as to its electrostatic potential\cite{Blasco03}. This is less attractive for a hole than that of an average Mn ion, whose charge at 1/3 doping is +3.3\cite{Alonso}.
 
\begin{figure}[tbp]
    \epsfxsize=8cm \epsfbox {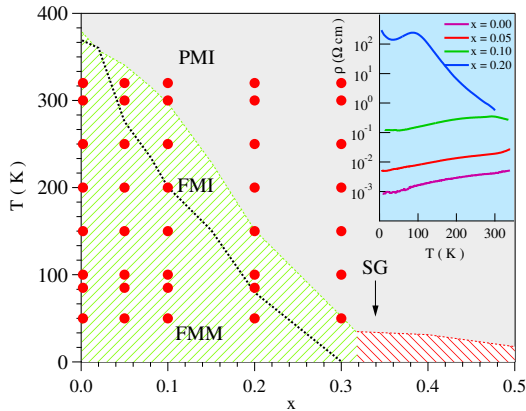}
\caption{Color online. Phase diagram of La$_{2/3}$Sr$_{1/3}$Mn$_{1-x}$Ga$_x$O$_3$ from Ref. \onlinecite{Blasco03}. The paramagnetic insulating (PMI), ferromagnetic metallic (FMM), ferromagnetic insulating (FMI), and spin-glass (SG) phases are shown. The red dots mark the reflectivity measurements performed in the present experiment. Inset: resistivity of the four La$_{2/3}$Sr$_{1/3}$Mn$_{1-x}$Ga$_x$O$_3$ pellets with a metallic phase, as prepared for the reflectivity measurements.} 
\label{diagram}
\end{figure}

The phase diagram of La$_{2/3}$Sr$_{1/3}$Mn$_{1-x}$Ga$_x$O$_3$ is shown in Fig.\ \ref{diagram} as it was reported in Ref. \onlinecite{Blasco03}.  In the Ga-free manganite a metallic FM phase (FMM) is established below $T_c$ = 380 K. By adding a small amount of Ga $T_c$ starts to decrease, until the FM phase disappears for $x$ = 0.30. Anomalies in the behavior of the resistivity vs. temperature have suggested\cite{Blasco03} the existence of an intermediate,  ferromagnetic insulating phase (FMI) between the paramagnetic insulator (PMI) and the FMM phase.  A spin-glass phase (SG) is also reported in the diagram beyond the $x,T$ range of the present investigation.

Infrared spectroscopy can independently observe the behavior of the free carriers and of the localized charges in different spectral regions. Then, it may greatly help to understand the effects of Ga substitution on the electrodynamics of a manganite. In the present study we have measured the reflectivity of five La$_{2/3}$Sr$_{1/3}$Mn$_{1-x}$Ga$_x$O$_3$  samples,  with increasing Ga content from 0.0 to 0.30, from 30 to 40000 cm$^{-1}$ and from 320 to 50 K. The ($x, T$) coordinates of the experimental points are marked by red dots on the phase diagram of Fig.\ \ref{diagram}. 
As far as we know, the only infrared spectrum of a Ga-substituted manganite was recently published\cite{Dubroka} in a Raman study of the phonon modes of La$_{2/3}$Sr$_{1/3}$Mn$_{1-x}$M$_x$O$_3$ , with M = Cr, Co, Cu, Sc, Zn, and Ga. Therein, the reflectivity of a sample with Ga $x$ = 0.08 was measured from 80 to 1200 cm$^{-1}$ to obtain its dc conductivity. Our study is aimed instead at observing the effect of increasing Ga substitution on the free-carrier absorption (Drude term), on the phonon spectrum, on the mid-infrared bands, and on the electronic absorption in the near infrared. We could thus follow, on the whole excitation spectrum, and under controlled  conditions, the destruction of a ferro-metallic phase and determine quantitatively the relationship between the magnetization and charge dynamics parameters like the plasma frequency, the effective mass, and the relaxation rate of the carriers.

\section{Experimental details}

To our knowledge, all studies on Ga-substituted manganites reported in the literature up to now concern polycrystalline materials. This is due to the difficulty to obtain a series of single crystals with different and controlled concentrations of Ga, which must also be stoichiometric in oxygen. In optical measurements, polished polycrystalline surfaces have been shown to provide an absolute mid-infrared reflectivity somewhat lower than in cleaved single crystals.\cite{Takenaka99,Takenaka00} This systematic error decreases as the metal is poorer and the reflectivity lower. For this reason, and being the present investigation aimed at comparing the behavior of samples with different Ga content prepared exactly in the same way, this unavoidable limitation is expected to affect marginally our results.

Here, La$_{2/3}$Sr$_{1/3}$Mn$_{1-x}$Ga$_x$O$_3$  samples with $x$ = 0.0, 0.05, 0.10, 0.20, and 0.30 were prepared in form of pellets and fully characterized as reported in Ref. \onlinecite{Blasco03}. The resistivity $\rho(T)$ of the pellets selected for the optical measurements was also measured - except for the $x$ = 0.30 insulator - by a standard four-wire procedure. The results are reported in the inset of Fig.\ \ref{diagram} and are in substantial agreement with the previous, more detailed measurements of Ref. \onlinecite{Blasco03} on different pellets with corresponding doping levels. 
The reflectivity $R(\omega)$ of the five samples was measured at nearly normal incidence after accurate polishing with sub-micron-thick powders. We used a rapid-scanning interferometer between 30 and 20000 cm$^{-1}$ and a grating monochromator from 16000 to 40000 cm$^{-1}$. The former data were taken by thermoregulating the samples within $\pm$ 2 K between 320 and 50 K, the latter ones at room temperature only. The reflectivity was then extrapolated both to zero and high frequency by use of accurate Drude-Lorentz fits. Afterwards, $\sigma(\omega)$ was extracted from $R(\omega)$ by standard Kramers-Kronig transformations.

\section{Results and discussion}

\begin{figure}[tbp]
	\epsfxsize=8cm \epsfbox {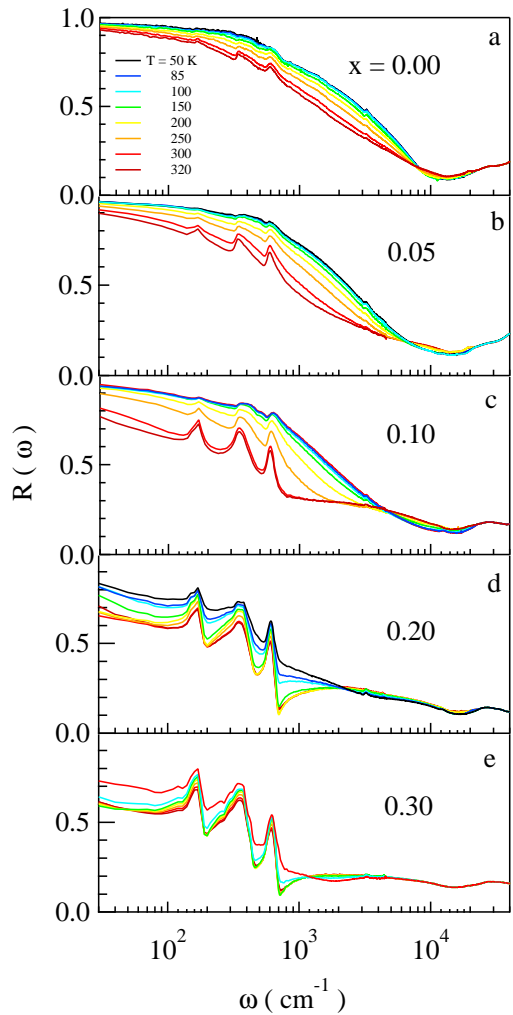}
\caption{Color online. Reflectivity $R(\omega)$ of the five La$_{2/3}$Sr$_{1/3}$Mn$_{1-x}$Ga$_x$O$_3$ samples, with $x$ increasing from top to bottom.} 
\label{refl}
\end{figure}

The measured reflectivity is shown in Fig.\ \ref{refl} at different temperatures for increasing Ga content, from top to bottom. In the infrared $R(\omega)$, in agreement with the behavior of $\rho(T)$ in the inset of Fig.\ \ref{diagram}, decreases  rapidly with the Ga content and, for a given $x$, for increasing temperature. Correspondingly, the three main infrared-active phonon lines - which are nearly shielded in a)-c) at low $T$ - show up. On the opposite side, a strong band in the visible (VIS) range is well evident in all samples. This band, with an edge just below 3 eV, is characteristic of manganites and is attributed to a charge-transfer transition between the O(2p) and the Mn(3d) orbitals\cite{Takenaka99,Okimoto}.

\begin{figure}[tbp]
	\epsfxsize=8cm \epsfbox {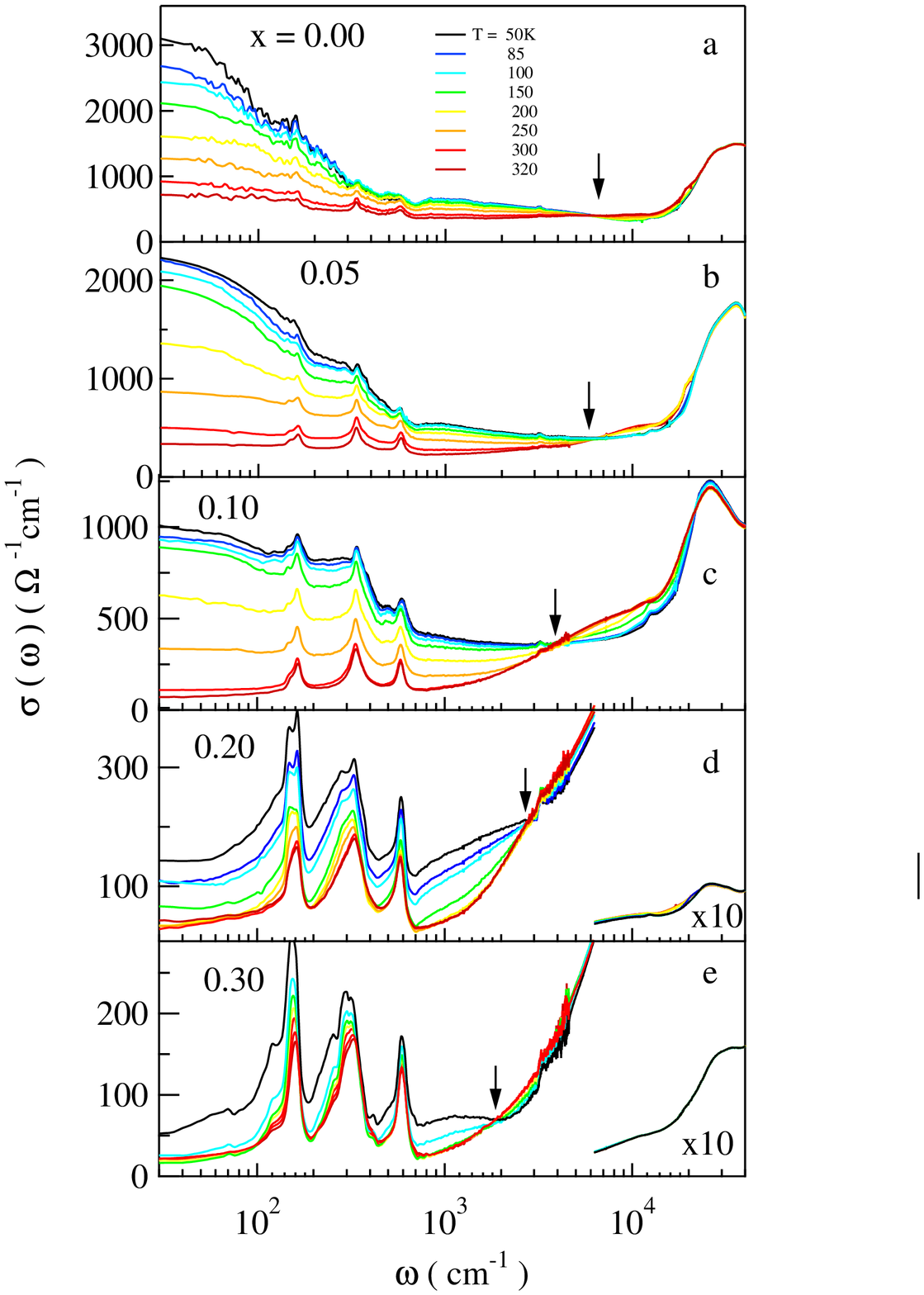}
\caption{Color online. Optical conductivity $\sigma (\omega)$ of the five La$_{2/3}$Sr$_{1/3}$Mn$_{1-x}$Ga$_x$O$_3$ samples, obtained from the reflectivity  of Fig. 3. The arrows mark the isosbestic points of the transfer of spectral weight (from high to low energy for decreasing $T$). In d) and e), the highest energy band has been reduced by a factor of 10.} 
\label{sigma}
\end{figure}

The real part $\sigma(\omega)$  of the optical conductivity extracted from the $R(\omega)$ of Fig.\ \ref{refl} is shown in Fig.\ \ref{sigma} for all samples. Therein, one can appreciate how the increasing Ga content (from top to bottom) turns gradually the infrared spectrum of the metallic manganite into that of an insulator, while the electronic bands at higher energy do not change significantly. The low-frequency values of 1/$\sigma(\omega)$ follow qualitatively both the $x,T$ dependence of $\rho$ in Fig.\ \ref{diagram}. Any quest for a closer correspondence between those two quantities would be meaningless in view of the grain-boundary effects\cite{Kim} which affect the dc current in pellets. A feature common to all samples is the transfer of spectral weight from high to low energy for decreasing $T$. This occurs around a very well defined isosbestic point, which softens as $x$ increases as shown by the arrows in the Figure. 

Both samples in a) and b) with Ga content 0.0 and 0.05, respectively, are metallic at all temperatures and their conductivity increases steadily upon cooling. As it will be discussed below, this effect is stronger than for an ordinary metal, where it is due to a simple reduction of the carrier scattering rate. A similar behavior is observed in c) for $x$ = 0.10, on a reduced conductivity scale. The shielding effect of the free carriers on the phonon lines here is also weaker than in a) and b). In d) the low-frequency conductivity is further reduced by more than a factor of 5 and the Drude continuum is nearly replaced by the phonon spectrum. This latter shows new lines in addition to the usual three infrared-active bands of doped manganites (see next Section). The transition to an insulating state induced by Ga is completed in the sample with $x$ = 0.30 in e). Therein however, a very weak dc conductivity can again be appreciated below 100 K. The phonon spectrum of this insulating sample, like that of $x$ = 0.20 in c), exhibits strong satellite lines. Moreover, a mid-infrared band centered at about 2000 cm$^{-1}$ appears at low $T$.  

The contributions to  $\sigma(\omega)$ in Fig.\ \ref{sigma} can be identified by a fitting procedure which provides consistent results from sample to sample.
To this purpose, one can use the usual Drude-Lorentz dielectric function

\begin{eqnarray}
\tilde\epsilon(\omega) = \epsilon_1(\omega) +i\epsilon_2(\omega) =  \epsilon_{\infty} - 
{\omega_{p}^2 \over {\omega^2 - i\omega \Gamma_{D}}} + \\ \nonumber
+ \sum_{j=1}^{n_{ph}} {S_{j}^2 \over {(\omega_{j}^2 - \omega^2) - i\omega\Gamma_{j}}}  
+ \sum_{k=1}^{n_{band}} {S_{k}^2 \over {(\omega_{k}^2 - \omega^2) - i\omega\Gamma_{k}}}\, , 
\label{D-L}
\end{eqnarray}

\noindent
with $\epsilon_2(\omega) = (4 \pi/\omega) \sigma(\omega)$. In the right side of Eq.\ \ref{D-L}, $\epsilon_{\infty}$ replaces all contributions at energies higher than the measuring range, while the second term is the Drude contribution with plasma frequency $\omega_{p}$ and relaxation rate $\Gamma_{D}$. The sum on $j$ includes all the lattice contributions with strength $S_{j}$ and width $\Gamma_{j}$, while the sum on $k$ describes the broad oscillators detected from the far infrared to the visible range. In the present case  $n_{band}$ = 4 and $k$ takes the values $FIR$ for an extra-phonon contribution in the far infrared, $MIR$ for the mid-infrared band, $NIR$ for a contribution observed in the near infrared, and $VIS$ for the band at 3 eV.

\begin{figure}[tbp]
	\epsfxsize=8cm \epsfbox {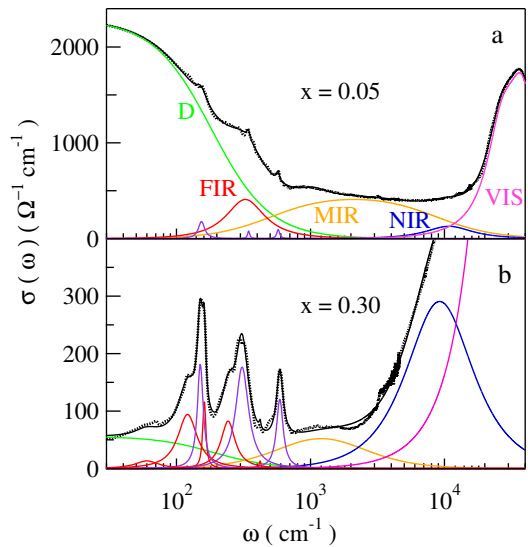}
\caption{Color online. Examples of Drude-Lorentz fits to the conductivity at 50 K of a metallic sample ($x$ = 0.05, a) and of an insulator ($x$ = 0.30, b). The black dotted line is the experimental curve, the black solid line the fit. Colored lines represent the Drude (D), far-infrared (FIR), mid-infrared (MIR), near-infrared (NIR), and visible (VIS) contributions (see text). The narrow peaks in the far infrared are phonon lines.} 
\label{fits}
\end{figure}

Examples of fits based on Eq.\ \ref{D-L} to the data of Fig.\ \ref{sigma} are shown in Fig.\ \ref{fits}. They illustrate two opposite cases at $T$ = 50 K: the $\sigma (\omega)$ of the metallic phase of $x$ = 0.05 (a) and that of the insulator $x$ = 0.30 (b). The dotted line is the experimental $\sigma (\omega)$, the black solid line the fitting curve, and the colored lines refer to the different contributions in Eq.\ \ref{D-L}. In the top panel one can easily distinguish the Drude (D), FIR, MIR, NIR, and VIS bands, in addition to weak phonon lines nearly shielded by the carrier background. These contributions are still present in the conductivity of the insulator (b), except for the FIR band. Therein, the latter contribution is resolved into satellite modes (red peaks) of the main phonon lines (in violet), as discussed below. In the following we will separately discuss those contributions to the conductivity for the five samples at all temperatures.

\subsubsection{Vibrational spectrum}

The decomposition of the $\sigma (\omega)$ for the insulating sample with $x$ = 0.30 (Fig.\ \ref{fits}-b) shows the usual three modes observed in most doped, pseudo-cubic, manganites: the external mode, here peaked at 152 cm$^{-1}$, the bending of the oxygen octahedra at 298 cm$^{-1}$, and their stretching at 590 cm$^{-1}$. 
These frequencies do not change appreciably with temperature, at variance with observations in other manganites. For example in La$_{0.875}$Sr$_{0.125}$MnO$_3$ all modes (the bending in particular) harden for decreasing temperature,\cite{Mayr} and are sensitive to the magnetic transitions. In view of the phase diagram of Fig. 1 it is not surprising, therefore, if the phonons of the  0.30 sample does not exhibit major shifts with $T$.

As already mentioned, the effect of Ga replacement manifests itself through satellite peaks which appear on both sides of the external and of the bending mode (Fig.\ \ref{fits}-b). These additional features are frequently observed in the insulating phases of doped oxides. They have been called either local modes or  InfraRed Active Vibrations (IRAV)\cite{Calvani} and are attributed to the vibrations of cells where the polaronic charges are self-trapped. Even when these ones are not numerous, IRAV's are observed due to the strong dipole moments associated with the vibrating polaron. This interpretation is supported by the fact that, when they are observed in oxides at low doping, a polaronic band also appears in the mid infrared which exhibits the same $T$ dependence: namely, the intensity of both the MIR band and the IRAV's rapidly increases as $T$ decreases and more and more charges are self-trapped. In this framework, IRAV's and MIR band correspond to transitions among the polaron internal states and to the free-carrier continuum, respectively. Both those spectral features are observed also in the present case, as shown in Fig.\ \ref{fits}-b) or in Fig.\ \ref{sigma}-e. 

The IRAV modes of Fig.\ \ref{fits}-b are very similar to those observed in the insulating cuprates at low doping \cite{Calvani96}.  Therein, they collapse into a broad band in the far infrared as doping increases and the metallic phase is approached. A similar effect is observed at $x$ = 0.05 at 50 K (Fig.\ \ref{fits}-a) where the introduction of a FIR band, peaked at about 380 cm$^{-1}$, is required by the fitting procedure. The FIR band becomes negligibly small at room temperature.

\subsubsection{Drude term}

\begin{figure}[tbp]
	\epsfxsize=8cm \epsfbox {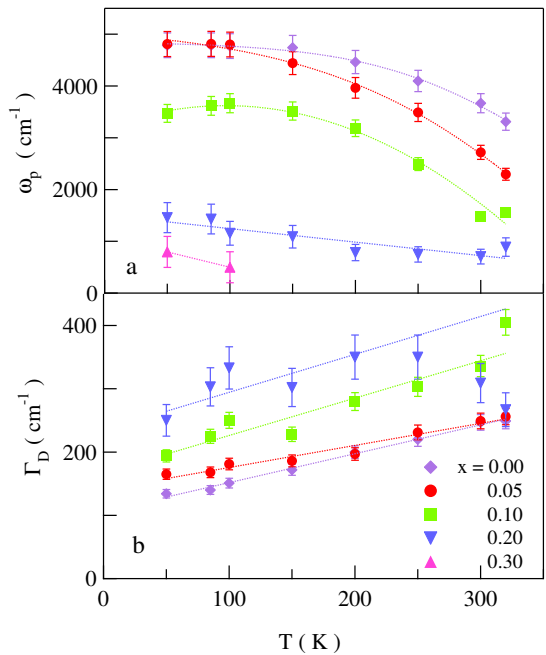}
\caption{Color online. Drude plasma frequency (a) and line width (b) for the five La$_{2/3}$Sr$_{1/3}$Mn$_{1-x}$Ga$_x$O$_3$  samples, as obtained from the fits to the optical conductivity of Fig. 4. At $x$ = 0.30, the fit includes a weak Drude term for $T \geq 100$ K. The lines are guides to the eye.}
\label{Drude} 
\end{figure}

The optical parameters which characterize the metallic state are the plasma frequency of the carriers $\omega_p$ and their scattering rate $\Gamma_D$. Those obtained by the fits to Eq.\ \ref{D-L} are plotted vs. temperature in Fig.\ \ref{Drude} for the four samples where the Drude contribution can be appreciated at all temperatures. An extremely small, but finite Drude term has been observed also in the sample with $x$ = 0.30 at the lowest temperature. 

One may notice in Fig.\ \ref{Drude}-a that $\omega_p$ decreases with the Ga content and exhibits a temperature dependence which is not observed in common metals. The following model may explain both these observations. 
In a simple one-band tight-binding model, the spectral weight of the Drude term, and then  $\omega_p^2$, is proportional to the hopping rate $t$ (see, \textit{e. g.}, Ref. \onlinecite{Ortolani})  between two generic sites $i$ and $j$. In turn, if $P_i^+ (P_i^-)$ is the probability that the ion spin at site $i$ is up (down), due to the Hund's rule
 
\begin{equation}
t \propto (P_i^+P_j^+ + P_i^-P_j^-) \, .
\label{t}
\end{equation}

\noindent
Let us call $M(x,T)$ the magnetization of a sample with Ga concentration $x$ at $T$ and $M_s = M(0,0) = 3.87 \mu_B$ (see Table I) the saturation magnetization in the pure FM phase of La$_{2/3}$Sr$_{1/3}$Mn$_{1-x}$Ga$_x$O$_3$  ($x$ = 0 and $T=0$). Of course, $M(x,T)$ = 0 in the PM phase. Therefore, $P_i^+ = [M_s + M(T)]/2M_s$ and $P_i^- = [M_s - M(T)]/2M_s$ for any $i$. Finally, including Eq. \ \ref{n},

\begin{equation}
\omega_p^2  \propto (M_s^2 + M^2(x,T))/(2M_s)^2 \, . 
\label{plasma2}
\end{equation}

\begin{figure}[tbp]
    \epsfxsize=8cm \epsfbox {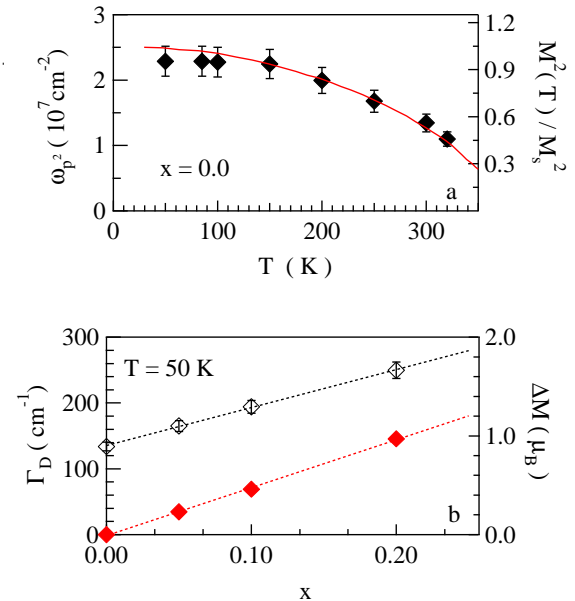}
\caption{Color online. a) Squared Drude plasma frequency vs. temperature (points, left scale) compared with the squared the magnetization  taken from Ref.\onlinecite{Uru} (red line, right scale). b) Both the Drude width $\Gamma_D$ at 50 K (black diamonds) and the difference $\Delta M = M(0,0) - M(x,0)$ between the saturation magnetization at $x$ = 0 and that at $x$ (red diamonds, right scale) are plotted vs. the Ga concentration $x$. The dashed lines are linear fits.}
\label{Gamma}
\end{figure}

\noindent
This prediction can be verified by comparing the behavior with temperature of $M(x,T)/M_s$ in La$_{2/3}$Sr$_{1/3}$Mn$_{1-x}$Ga$_x$O$_3$  with the $\omega_p (T)$ of Fig.\ \ref{Drude}. Presently, the former datum is available only for\cite{Uru} $x$ = 0. Both the $\omega_p^2 (T)$ here measured in the Ga-free sample and $[M(0,T)/M_s]^2$ are plotted vs. $T$ in Fig.\ \ref{Gamma}-a, where Eq.\ \ref{plasma2} is well verified. One can conclude that the above model provides a good picture of the interplay between charge transfer and magnetic order in a La-Sr manganite.

Figure\ \ref{Gamma}-b shows instead the Drude width $\Gamma_D$ as a function of Ga doping at $T$ = 50 K. As $\Gamma_D$ is a scattering probability, it can be written as a sum of independent processes as  

\begin{equation}
\Gamma_D (x,T) = \Gamma_{ph}(T) + \Gamma_{str}(x) + \Gamma_{mag}(x,T) \, .
\label{Gamma_D}
\end{equation}

\noindent
Therein, the first term on the right is due to phonon scattering, which is assumed to be independent of $x$ in view of the low perturbation of Ga on the lattice dynamics. The second one is due to the structural disorder in the Mn-O planes caused by Ga impurities. At not too high $x$, it can be assumed to be proportional to their density: $\Gamma_{str}(x) \propto x$. The third term is due to the magnetic dilution due to Ga substitution. At $T << T_c$ (here, at 50 K) the resulting increase in the scattering rate can be measured by the difference
$\Delta M = M(0,0) - M(x,0)$ between the saturation magnetization of the Ga-free manganite and that of the sample with Ga concentration $x$. As a first approximation we can assume $\Gamma_{mag}(x) \propto \Delta M $. In turn $\Delta M$, as obtained from the values of $M(x,0)$ in Table I, is shown in  Fig. \ \ref{Gamma}-b to depend linearly on $x$ (red diamonds, right scale)for $x \leq 0.20$. Therefore, from Eq.\ \ref{Gamma_D} at low and constant $T$, 
\begin{equation}
\Gamma_D (x, 50 K) = \Gamma_0 + \Gamma_{str}(x) + \Gamma_{mag}(x) = \Gamma_0 + const. \cdot x \,. 
\label{Gamma_0}
\end{equation}

\noindent
This simple relation describes well the observations, as shown in the same Fig. \ \ref{Gamma}-b (black diamonds, left scale). 

From the present infrared data one can also find how the effective mass of the carriers increases for increasing Ga content. Indeed, in the Drude model, the plasma frequency is 

\begin{equation}
\omega_p^2 = {1\over 4 \pi^2}{ne^2 \over m^\ast} \, ,
\label{omega_p}
\end{equation}

\noindent
where $n$ is the number of carriers per cm$^3$ and $m^\ast$ is the effective mass.
Let us first evaluate $n$. In manganites, the transport is regulated by the combined role of Mn$^{+3}$ and Mn$^{+4}$ ions. A generic manganite A$_{1-y}$B$_{y}$MnO$_{3}$  with $y$ acceptors per formula unit, in the simplest approach to the double-exchange charge transfer, will be insulating when either $y$ = 0 (all ions are Mn$^{+3}$) or $y$ = 1 (all ions are Mn$^{+4}$). This is for example the case of La$_{1-y}$Ca$_{y}$MnO$_{3}$, even if its phase diagram is not symmetric with respect to $y$ due to magnetic, Jahn-Teller, and other effects. In the above simple approximation, the number of available carriers is $n \propto y(1-y)$. 
In the present case, where each Ga$^{+3}$ suppresses one Mn$^{+3}$ while the Mn$^{+4}$ concentration remains unaltered, one can assume for the carrier concentration 

\begin{equation}
n = y(1-y-x) /V_c \, ,
\label{n}
\end{equation}

\noindent
where $V_c$ is the formula-unit cell volume. From Eq.\ \ref{n} one obtains $n = 0.33(0.67-x) /V_c$. Then, by solving Eq.\ \ref{omega_p} with respect to  $m^\ast$ one obtains the values reported in Table I for the ratio $m^\ast/m_e$, where $m_e$ is the mass of a free electron.

\begin{table}[tbp]
\caption{Parameters of the model employed to account for the Drude term and the MIR band in La$_{2/3}$Sr$_{1/3}$Mn$_{1-x}$Ga$_x$O$_3$  and comparison with the observations at 50 K. The table reports the volume $V_c$ of the unit cell, the $M(x$, 0) values reported at 5 K in Ref.\ \onlinecite{Blasco03}, the number of carriers available per unit cell, the effective mass of the carriers estimated from $\omega_p$, the intensity of the MIR band, as calculated  by Eq.\ \ref{mir-intensity2}, and that measured in the present experiment. }

\label{Table I}
\begin{ruledtabular}
\begin{tabular}{ccccccc}

$x$   & $V_c$  & $M(x$,0) & $n/cell$   & $m^\ast/m_e$ & $ S^2_{MIR}$  & $S^2_{MIR}$ \\
	& (nm$^3$) & ($\mu_B$) &		&		&	(calc.) &	(exp.) \\
\colrule
0.00  & 0.34999  &   3.87	   &   0.221  	&   1.6 $\pm$ 0.3	&   1     	& 1	  \\
0.05  & 0.34970  &   3.64   &   0.206  	&   1.6 $\pm$ 0.3	&   0.88  & 0.78 \\
0.10  & 0.34937  &   3.41   &   0.189 	&   2.9 $\pm$ 0.6	&   0.76  & 0.73 \\
0.20  & 0.34885  &   2.90   &   0.157	&    13 $\pm$ 3		&   0.55  & 0.47 \\
0.30  & 0.34826  &   1.76   &   0.123	&    35 $\pm$ 7	 	&   0.34  & 0.32 \\

\end{tabular}
\end{ruledtabular}
\end{table}

Table I points out the rapid increase in the carrier effective mass, and then in the charge localization, with the gallium concentration $x$. In samples with $x$ = 0.20 and 0.30, $m^\ast$ reaches values typical of small polarons, namely of charges which are assumed to be self-trapped within a single cell. In Table I, the effective mass in the Ga-free La-Sr manganite is somewhat smaller than those reported (at room temperature only) for a single crystal of La$_{0.7}$Sr$_{0.3}$MnO$_3$ in Ref. \onlinecite{TakenakaJPSJ} and for a thin film of the same material in Ref. \onlinecite{Simpson}. However, in both cases the authors assume a pure Drude term, with no MIR band, and therefore a larger number of free carriers.

\subsubsection{Mid- and near-infrared bands}

The Drude-Lorentz fit to $\sigma(\omega)$ allows one to identify in all spectra two bands. The first one, the MIR band, shows at all doping a characteristic behavior with temperature: its peak frequency softens as the insulator-to-metal (IM) transition is approached, to remain stable at about 2000 cm$^{-1}$ in the whole metallic phase. The other one, the NIR band, is instead centered at about 10000 cm$^{-1}$ in all samples. 

\begin{figure}[tbp]
    \epsfxsize=8cm \epsfbox {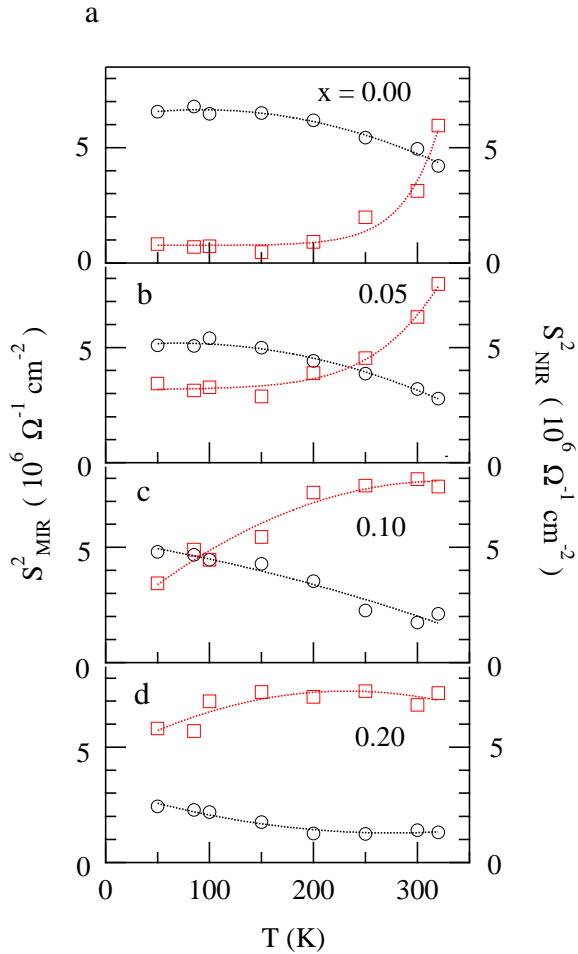}
\caption{Color online. Intensity vs. temperature of the bands MIR (in black, left scale) and NIR (in red, right scale) in La$_{2/3}$Sr$_{1/3}$Mn$_{1-x}$Ga$_x$O$_3$ with $x$ varying from 0.0 to 0.20. The lines are guides to the eye.}
\label{mir-nir}
\end{figure}

The band intensities $S_{MIR}^2$ and $S_{NIR}^2$ in Eq.\ \ref{D-L} are plotted in Fig.\ \ref{mir-nir} vs $T$. At low Ga doping ($x$ = 0.05 and 0.10), one may appreciate how the insulating-to-metal transition is triggered by a transfer of spectral weight from the NIR to the MIR band. This occurs around an isosbestic point $\omega_{iso}$ which can be easily identified in the reflectivity curves of Fig.\ \ref{refl}. $\omega_{iso}$ decreases from 7800 cm$^{-1}$  for $x$ = 0, to 6500 cm$^{-1}$ for $x$ = 0.05, 4500 cm$^{-1}$ for $x$ = 0.10, 2100 cm$^{-1}$ for $x$ = 0.20, and 1150 cm$^{-1}$ for $x$ = 0.30.

A MIR band has been observed in virtually all the conducting manganites. Even if it was proposed\cite{Takenaka99,Takenaka00} that it may be an artifact related to the use of polycrystalline samples, a bare Drude term cannot explain the optical conductivity even in the best metallic single crystals \cite{nota}. In the framework of an ionic model for the Mn-O planes, the MIR band has been attributed\cite{Jung,Jung99,Nucara06} to the polaronic charge-transfer Mn$^{+3}$ - Mn$^{+4}$. Such hopping indeed implies an adiabatic transition between the Jahn-Teller distorted Mn$^{+3}$ $e_{g1}$ state and the undistorted Mn$^{+4}$ $e_g$ state. This interpretation of the MIR band is confirmed by Fig.\ \ref{mir-nir}, where the MIR intensity is sensitive to the insulator-to-metal transition. Moreover, as it will be shown below, such intensity is governed by the same magnetic mechanism (Eq.\ \ref{plasma2}) which governs the plasma frequency of the Drude term. Indeed, according to theoretical models\cite{Perroni}, the mid-infrared and the Drude contributions are the incoherent and coherent parts, respectively, of a polaronic $\sigma (\omega)$.

If the MIR band is due to photon-promoted hopping from site $i$ to site $j$, its intensity will depend on the magnetization as in Eq.\ \ref{plasma2} for the intensity of the Drude term, and in Ref. \onlinecite{Jung99} for the MIR band of La$_{7/8}$Sr$_{1/8}$MnO$_3$.  After including also the number of charges available for hopping, from Eq.\ \ref{n}, one has:

\begin{equation}
S^2_{MIR} \propto y(1-x-y)\cdot (M_s^2 + M^2 (T))/(2M_s)^2 \,.
\label{mir-intensity2}
\end{equation}

\noindent
Using the values $M(x,$ 0) measured\cite{Blasco03}  at 5 K,  and reported in Table I, one obtains from Eq.\ \ref{mir-intensity2} the results shown in the same Table. Therein, the intensity $S^2_{MIR}$ of the MIR band at $x$ = 0.0 has been taken equal to 1. In the last column of the same Table, the intensity of the MIR band observed here at the same $x$ and 50 K is reported for comparison. 

Table I shows that a model based on photon-assisted hopping well accounts for the decrease of the MIR band intensity with Ga doping. This suggests that both effects of Ga, namely the reduction in the number of carriers and the weakening of the FM order, well describe the evolution of the mid-infrared spectrum up to the collapse of the FM metallic phase. 

Unlike the MIR band, the NIR band at about 10000 cm$^{-1}$ is most likely purely electronic. It is attributed to the transitions $e_{g1} - e_{g2}$, either on the same site or between two adjacent Mn$^{+3}$ sites\cite{Loidl}. It is then reasonable that it transfers part of its spectral weight to the MIR band (see Fig.\ \ref{mir-nir}) when the transition to the FM phase switches on the charge hopping. The $NIR$ absolute intensity should scale with $x$, but a proportionality is observed between 0.05 and 0.10 only.

\section{Conclusion}

The present experiment was aimed at first studying, by infrared spectroscopy, the metal-to-insulator transition induced by the substitution of Ga for Mn in a metallic manganite like La$_{2/3}$Sr$_{1/3}$MnO$_3$. Indeed, Ga impurities dilute both mechanisms involved in the intersite charge transfer, \textit{i. e.}, ferromagnetism and Jahn-Teller distortion, without major effects on the lattice properties. 

The experiment provided several interesting results. First, the intensity of the Drude term $\omega_p^2$ and that of the mid-infrared band $S^2_{MIR}$ change both with temperature and with the Ga concentration $x$. The behavior of $\omega_p^2$ with $T$ at a given $x$, and that of the MIR band with $x$ at low $T$, are shown to be directly governed by the magnetization, through a simple quadratic dependence. This finding also confirms that, in the spectrum of a metallic manganite, those absorption features are strictly related with each other. As it has been suggested theoretically\cite{Perroni}, they represent the coherent and the incoherent part of a polaronic optical conductivity, respectively. The $T$-dependence of $S^2_{MIR}$ is mainly determined by a transfer of spectral weight from a band in the near infrared. This occurs around a very well defined isosbestic point whose frequency decreases for increasing $x$. One should recall that a similar softening of the spectral weight for $T \to$ 0 is observed in many systems characterized by strong electron-electron correlations\cite{Lupi92}. 

Secondly, a linear increase with $x$ in the low-temperature Drude linewidth $\Gamma_D$ was observed, and explained by a simple model which takes into account both the structural and the magnetic disorder induced by Ga impurities. 
Finally, the effective mass $m^\ast$ of the carriers, here monitored directly by the plasma frequency $\omega_p$, was found to increase much more than linearly with $x$: $m^\ast $ changes from 1.6 bare-electron masses in the Ga-free sample, to about 35 electron masses as the system eventually turns into an insulator. Such value indicates that the carriers, at $x$ = 0.30, are small polarons. Meanwhile, in the far infrared aside the phonon lines, satellite bands do appear which can be attributed to local modes of the cells distorted by the localized charges.

In summary, the present optical spectra allowed us to monitor directly, at different energy scales, how an increasing dilution of the ferromagnetic order reflects in the charge dynamics of a metallic manganite, until it is eventually turned into an insulator at all temperatures.

\acknowledgments

The authors are indebted to Javier Blasco for supplying the well characterized samples here studied.



\begin{thebibliography}{}
\bibitem{nota} 
We have unsuccessfully tried to fit the $R(\omega)$ of a single crystal of La$_{2/3}$Sr$_{1/3}$MnO$_3$ in Ref. \onlinecite{TakenakaJPSJ} by using a single Drude term, plus the usual phonon lines and high-energy bands. A very good fit was instead obtained by adding a strong and broad mid-infrared band peaked at 2400 cm$^{-1}$.
\end{thebibliography}
\end{document}